\documentclass[10pt, conference, compsocconf]{IEEEtran}
\usepackage[font=footnotesize,skip=\baselineskip]{caption}
\usepackage{mathptmx} 
\usepackage{comment}
\usepackage{fancyhdr}
\usepackage[normalem]{ulem}
\usepackage[hyphens]{url}
\usepackage[sort,nocompress]{cite}
\usepackage[final]{microtype}
\usepackage{flushend}
\usepackage[bookmarks=true,breaklinks=true,letterpaper=true,colorlinks,linkcolor=black,citecolor=blue,urlcolor=black]{hyperref}
\usepackage{fancyhdr}
\usepackage[displaymath,mathlines,pagewise]{lineno}
\usepackage{color,soul}
\definecolor{lightgreen}{RGB}{195, 233, 211}
\definecolor{orange}{RGB}{207,83,0}
\sethlcolor{lightgreen}
\usepackage[ruled,vlined,linesnumbered]{algorithm2e}
\usepackage{algorithmic}
\usepackage{cite}
\usepackage{mathptmx}
\usepackage{setspace}
\usepackage{mathdesign}
\usepackage{adjustbox}
\usepackage{graphicx,dblfloatfix}
\usepackage[labelfont=bf]{caption}
\usepackage{epstopdf}
\usepackage[nopar]{lipsum}
\usepackage{graphicx}
\usepackage{mathtools}
\usepackage{amssymb,amsmath}
\usepackage[numbers,compress]{natbib}
\usepackage{footnote}
\makesavenoteenv{tabular}
\makesavenoteenv{table}
\usepackage{tablefootnote}
\usepackage{color}

\usepackage {pifont}
\usepackage{array}
\usepackage{wasysym}
\usepackage{multirow}
\usepackage{xfrac}
\usepackage[utf8]{inputenc}
\usepackage{etoolbox}
\usepackage{framed}
\definecolor{shadecolor}{RGB}{255,255,0}
\newbool{inccomment}
\setbool{inccomment}{true}
\newcommand{\sms}[1]{\ifbool{inccomment}{{\color{red}#1}}{}}
\newcommand{\akj}[1]{\ifbool{inccomment}{{\color{orange}#1}}{}}
\newcommand{\rami}[1]{\ifbool{inccomment}{{\color{black}#1}}{}}
\newcommand{\ignore}[1]{}
\usepackage[pass]{geometry}
\usepackage{fancyhdr}
\usepackage[normalem]{ulem}
\usepackage[hyphens]{url}
\usepackage{hyperref}
\usepackage{natbib}
\ifCLASSINFOpdf
\else
\fi
\begin{document}
\fancypagestyle{firstpage}{
\fancyhf{}
\setlength{\headheight}{50pt}
\renewcommand{\headrulewidth}{0pt}
\fancyhead[C]{\normalsize{
\textbf{}}}
\pagenumbering{arabic}
}

\newcommand{\superscript}[1]{\ensuremath{^{\textrm{#1}}}}
\def\sharedaffiliation{\end{tabular}\newline\begin{tabular}{c}}
\def\wg{\superscript{\dag}}
\def\wu{\superscript{\S}}
\title{\huge Enabling Fine-Grain Restricted Coset Coding Through Word-Level 
Compression for PCM\vspace{-0.5em}}

\author{\IEEEauthorblockN{Seyed Mohammad Seyedzadeh\wg, Alex K. Jones\wu, Rami Melhem\wg\\}
\IEEEauthorblockA{Computer Science Department\wg, Electrical and Computer Engineering Department\wu\\
University of Pittsburgh\\
\normalsize seyedzadeh@cs.pitt.edu, akjones@pitt.edu, melhem@cs.pitt.edu\vspace{-0.5em}
}
}
\setstretch{0.845}
\thispagestyle{firstpage}
\pagestyle{plain}
\makeatletter
\makeatother
\maketitle
\textit{Phase change memory (PCM) has recently emerged as a promising technology to meet the fast growing demand for large capacity memory in computer systems, replacing DRAM that is impeded by physical limitations. Multi-level cell (MLC) PCM offers high density with low per-byte fabrication cost. However, despite many advantages, such as scalability and low leakage, the energy for programming intermediate states is considerably larger than programing single-level cell PCM.
In this paper, we study encoding techniques to reduce write energy for MLC PCM when the encoding granularity is lowered below the typical cache line size. We observe that encoding data blocks at small granularity to reduce write energy actually increases the write energy because of the auxiliary encoding bits. We mitigate this adverse effect by 1) designing suitable codeword mappings that use fewer auxiliary bits and 2) proposing a new Word-Level Compression (WLC) which compresses more than 91\% of the memory lines and provides enough room to store the auxiliary data using a novel \textit{restricted} coset encoding applied at small data block granularities.}

\textit{Experimental results show that the proposed encoding at 16-bit data granularity reduces the write energy by 39\%, on average, versus the leading encoding approach for write energy reduction. Furthermore, it improves endurance by 20\% and is more reliable than the leading approach. Hardware synthesis evaluation shows that the proposed encoding can be implemented on-chip with only a nominal area overhead.}

\section{\textbf{Introduction}}
Scaling down process technology enables higher capacity main memories by increasing the density of DRAM devices. Unfortunately, this trend jeopardizes current DRAM main memory designs, especially when scaling beyond 20nm technology, because of fundamental obstacles related to high power consumption and process variation problems~\cite{zhang2017balancing,seyedzadeh2017counter,seyedzadehmitigating}. Among several alternative technology candidates, Phase-Change Memory (PCM) is emerging as promising due to its desirable characteristics in terms of scalability, low access latency and negligible standby power~\cite{lee2009architecting,raoux2008phase}. 

To store data in PCM cells, some PCM prototypes consider only two states per storage element, SET and RESET, to produce a single level cell~(SLC). However, the large resistance range between SET and RESET has motivated a split of that range into different states to enable multi-level cells. Contrary to SLC that can represent a logic `1' or `0', a MLC requires multiple programming write pulses to adjust the cell resistance to one of many predetermined ranges. For this reason, MLC PCM usually adopts an iterative program-and-verify (P\&V) technique~\cite{cabrini2009voltage,jiang2012improving,bedeschi2009bipolar} to achieve precise resistance control.

The P\&V technique first resets the cell and then iteratively uses partial SET pulses and checks whether the predetermined resistance range is reached. There are two main disadvantages to this type of write programming. The first is that resetting a cell before iteratively programming it negatively impacts the cell endurance~\cite{cho2009flip}. The second is that it increases write energy in the system~\cite{dong2011adams}. Specifically, the write energy of MLC PCM has been reported to be 10 times more than that of the SLC PCM~\cite{wang2011energy,jadidi2017exploring}. Moreover, when scaling below 20nm technology, the high energy needed to write a cell may disturb neighboring cells that are idle--i.e., not changed by the write. This phenomenon is known as write disturbance in MLC PCM~\cite{jiang2014mitigating}. 

To reduce the write energy in PCM, improve endurance or alleviate write disturbance, we will discuss in Section~\ref{related} several already proposed solutions that are based on encoding memory lines. Our goal in this work is to reduce write energy in PCM with 4-level cells, which we can abstract as 2-bit \emph{symbols}, by designing a simple and effective encoding architecture that lowers the encoding granularity while still incurring very low encoding overhead. Reducing the write energy, however, should not adversely affect the endurance of the PCM or its susceptibility to write disturbance.
Based on a comprehensive characterization, we base our contribution on five observations related to the effect of encoding on the write energy, endurance and write disturbance errors.

First, we observe that when the encoding data block granularity is reduced, the number of auxiliary symbols increases and the write energy resulting from auxiliary symbols may neutralize or exceed the write energy savings resulting from encoding the data. 
Second, we observe that symbol frequency distribution in real workloads is not random, and hence the choice of the alternative code words used for encoding should take into consideration both the write energy of each symbol as well as its frequency probability in real workloads.
Third, we observe that although current memory line compression techniques can achieve a large compression ratio, it can only compress a small fraction of the memory lines. A lighter compression technique which achieves a smaller compression ratio but successfully compresses more lines is more suitable for accommodating the auxiliary symbols needed for an encoding.
Fourth, we observe that words in a memory line have similar characteristics, and hence, when applying code mappings to different subblocks of a memory line, it is possible to use the same mappings for all the subblocks in the line. This restricted encoding has a minimal effect on the write energy but reduces the number of auxiliary symbols. 
Fifth, we observe that an encoding that reduces the write energy also reduces the number of written cells as well as the number of written symbols causing write disturbance. Hence, reducing the write energy through fine encoding granularity 
also improves cell endurance and reduces the probability of write disturbance.

Based on our observations, we propose \textbf{WLCRC}, a \textbf{W}ord-\textbf{L}evel
\textbf{C}ompression \textbf{R}estricted \textbf{C}oset coding architecture that achieves the best write energy savings at 16-bit encoding granularity. This work makes the following contributions:
\begin{itemize}
\item We revisit codeword mappings that map high energy states to low energy states and show that the selection of suitable codeword mappings that take into consideration the bit patterns frequently occurring in data directly impacts the write energy.
\item We characterize the write energy for biased and random workloads for coset coding techniques when the data block granularity is changed between 512 and 8 bits. Our analysis reveals the five observations mentioned above that are useful for developing new mechanisms to improve write energy, endurance and reliability.
\item We design a Word Level Compression~(WLC) technique that compresses more than 90\% of the memory lines while using very simple compression/decompression logic.
\item We propose a new and low overhead fine-grained `restricted coset encoding' that can be integrated with WLC. Through light compression, WLC provides enough room within memory lines for storing auxiliary symbols of fine granularity restricted coset encoding.
\end{itemize}
\section{\textbf{Background and Motivation}}
In this section, we first provide some necessary background on writing data in MLC PCM. Next, we discuss write energy, endurance and write disturbance errors. Finally, we briefly review related work and illustrate the motivation of our paper.\vspace{-0.29em}

\subsection{\textbf{PCM}}
A single PCM cell is programed by switching the chalcogenide material between a high resistance amorphous state (RESET) and a low resistance crystalline state~(SET) through the application of a programming current. The cell in the SET state requires a high intense programming current to change its status to the RESET state. This current is considerably more than the current required for switching the cell from RESET to SET. Studies showed that the loss of cell endurance is directly correlated to the high programming current~\cite{zhang2009characterizing,kim2005reliability}.

The large resistance contrast between the SET and RESET states enables the exploitation of partially crystallized states to store more than one bit per cell (Multi-level cell). In current MLC PCM, the resistance range between the RESET and the SET states is split into four regions that represent the logic values `00', `01', `10', and `11'. Since it is impractical to precisely program MLC cells through a single pulse, industrial prototypes and academic research resort to an iterative program-and-verify~(P\&V) policy~\cite{nirschl2007write,pantazi2009multilevel}. Unfortunately, P\&V programming increases the write energy by a factors reaching 10$\times$ that of programming a SLC~\cite{wang2011energy}.

The high heat resulting from resetting a cell, particularly in MLC PCM, may disturb neighboring \textit{idle} cells that are not being programmed. Specifically, the generated heat reduces the resistance of the idle cells and may unintentionally put them in the SET state. This reliability bottleneck increases when memory cells are scaled below 20nm technology where cell-to-cell distance decreases considerably~\cite{kim2011current,ahn2011reliability,jiang2014mitigating}. Note that if all cells are updated in a write operation, there are no idle cells and consequently no write disturbance. However, because the endurance of a PCM cell is determined by the number of writes to that cell, Differential Write ~\cite{zhou2009durable} is used and only cells which are actually changed are written.
\vspace{-0.5em}

\subsection{\textbf{Related Work}}\label{related}
Several techniques have been proposed to confront high write energy~\cite{imani2017exploring,imani2016low,imani2016loww,seyedzadeh2015pres,seyedzadeh2016improving,seyedzadeh2016leveraging}, endurance and write disturbance problems in PCM. The key concept behind all of them is to reduce the number of state changes (write operations) that are costly in terms of energy, endurance and write disturbance. Data encoding is a common solution that effectively reduces the number of costly cell programming operations. For example, Flip-N-Write~\cite{cho2009flip} was proposed for SLC PCM to reduce the number of written cells in the memory. 
To improve the lifetime of SLCs, FlipMin~\cite{jacobvitz2013coset} was proposed based on the concept of coset encoding~\cite{forney1988coset}. The basic idea of FlipMin is to perform a one-to-one mapping from the data block to a coset of code word candidates. Then, the code word candidate that optimizes the lifetime is selected to be written in the memory. The initial coset candidates are build by the dual code of a (72,64) Hamming generator matrix. Since the initial coset candidates are essentially random binary vectors, FlipMin is most effective for workloads operating on random data~\cite{seyedzadeh2016improving}. 

To reduce write energy in MLC PCM and achieve low encoding overhead, an encoding that uses six coset candidates has been proposed in~\cite{wang2011energy} with the goal of mapping the two high energy states to the two low energy states. 

To mitigate write disturbance errors in SLC and MLC wordlines, a 3-to-4-bit encoding, DIN, was proposed in~\cite{jiang2014mitigating} and was integrated with a 20-bit BCH code to correct any two write disturbance errors in a verification step. To make room for the extended code words,
DIN uses memory line compression to compresses a 512-bit line to 369-bits.
However, because of the required large compression ratio, DIN is only able to compress and encode 30\% of the memory lines.
\vspace{-0.25em}
\subsection{\textbf{Motivation}}
Using simulation (see Section \ref{sec:setting}), we measured the write energy when the coset encoding proposed in~\cite{wang2011energy} is used along with differential write.
Figure~\ref{motivation}(a) shows the results for 200 million 512-bit random data lines when the encoding granularity ranges between 8 and 512 bits. At a given granularity of $x$-bits, each $x$-bits data block is separately encoded using one of six coset candidates at the cost of adding 3-bits (two auxiliary symbols in MLC PCM) to identify the candidate used.
The figure breaks down the write energy into the energy to write the data and the auxiliary symbols. It shows that when the encoding granularity decreases, the write energy and its dominant component, the data symbol energy, decreases. On the other hand, the auxiliary symbol energy gradually increases and reaches its maximum at the granularity of 8-bits. 

We also performed a similar study on real workloads to investigate the relationship between the two components of the write energy. Figure~\ref{motivation}(b) shows that the energy for the real (biased) workloads is considerably smaller than the random workload case, which is due to data locality. However the trend with varying granularity is the same for both workloads. The main reason for energy reduction at small data block granularity is the flexibility of encoding smaller data blocks independently.
Unfortunately, this benefit comes at the expense of a high space overhead needed for storing the auxiliary symbols. This overhead reaches 25\% at 8-bit granularity (two auxiliary symbols for each four data symbols).

\textbf{Our goal} in this paper is to take advantage of fine-grain encoding granularity while reducing the overhead of auxiliary symbols and introducing a light weight compression that provides enough space in the memory line to store auxiliary symbols.
\vspace{-0.8em}

\section{\textbf{Revisiting Coset Candidates}}\label{coset}
A 4-level cell can be programmed to any one of four resistance states. We denote these states by S1, S2, S3 and S4 and we assume that the states are numbered in the order implied by the energy needed to bring a cell to that particular state, with S1 requiring the least energy and S4 requiring the most energy (see Table \ref{four_coset_candidates}). Specifically, programming into S1 is done using a RESET pulse, while programming it into S2 is done using a SET pulse, which consumes more energy. Programming into S3 and S4 is done through iterative partial SET pulses~\cite{joshi2011mercury}. Note that to reach S2, S3, S4, the cell must be first reset before applying the SET pulses.
Every two consecutive bits in a memory line are stored in one cell. Hence, an encoding is a particular mapping of the four symbols, `00', `01', `10' and `11' into the four cell states. We assume that the default mapping of the four symbols `00', `10', `11', and `01' is to the states S1, S2, S3 and S4, respectively~\cite{jiang2014mitigating}.

\begin{figure}[!t]
\centering
\includegraphics[width=0.31\textheight]{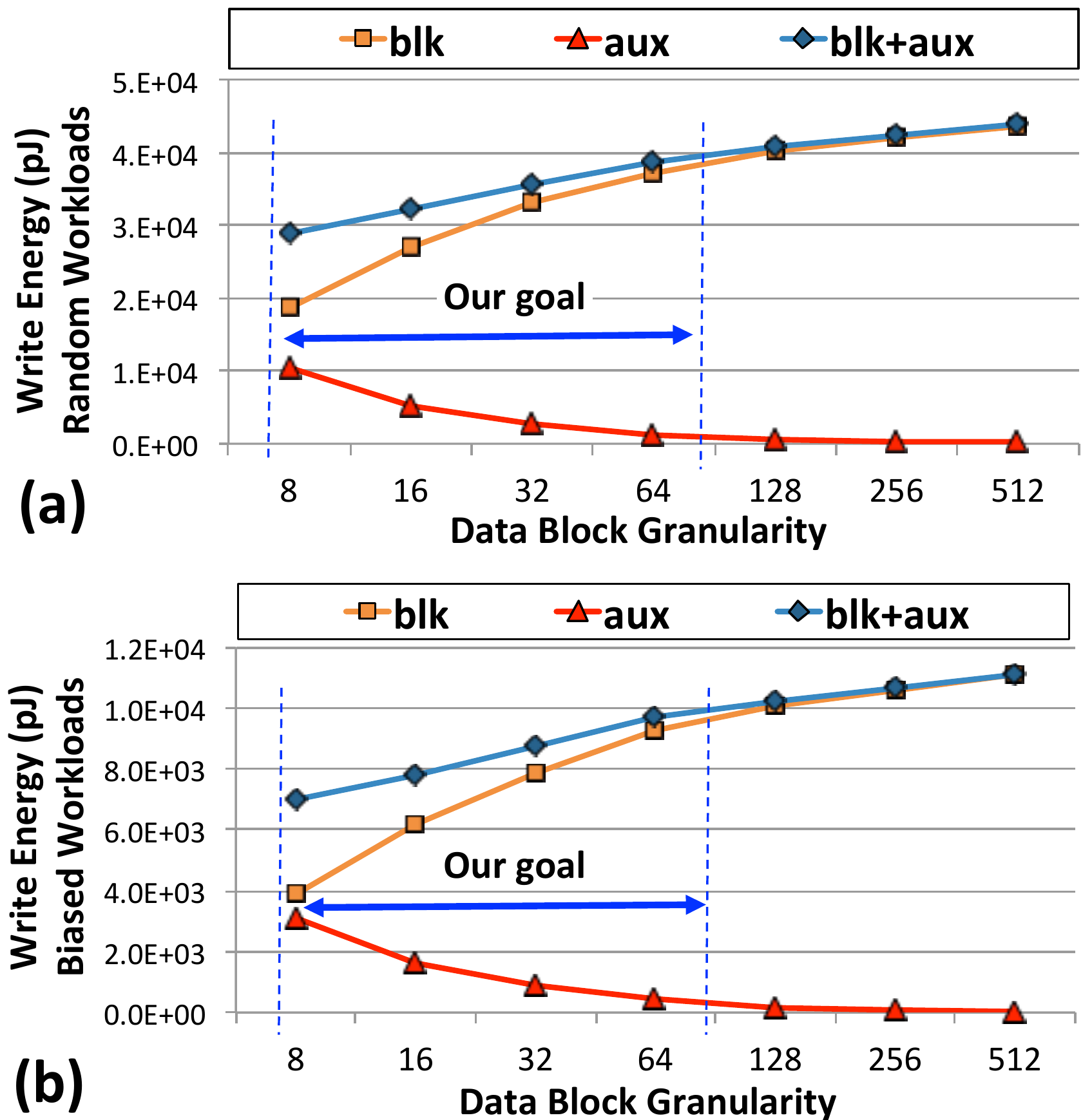}
\caption{ Write energy analysis for (a) random workloads; (b) biased workloads (SPEC2006 and PARSEC benchmarks).\vspace{-1.75em}
}
\label{motivation}
\end{figure}

\begin{figure*}[!t]
\centering
\includegraphics[width=0.725\textheight]{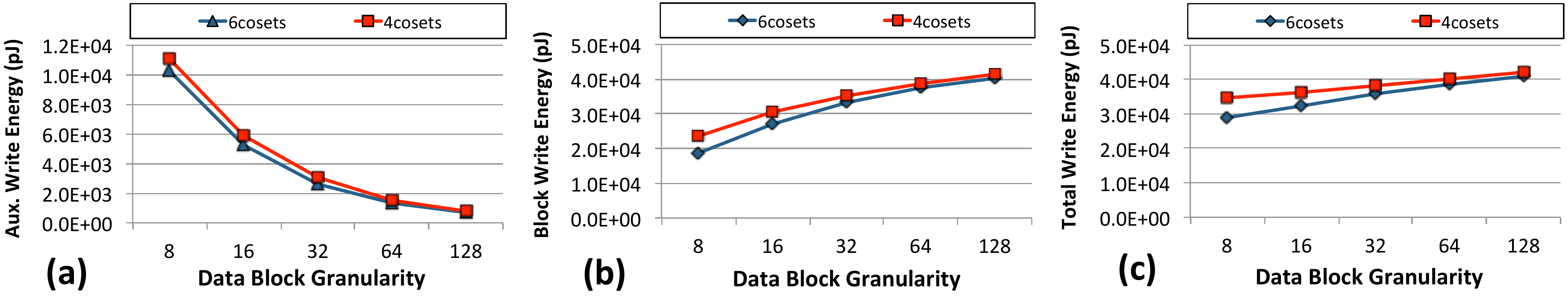}\vspace{-0.5em}
\caption{ Write energy analysis for (a) auxiliary symbols, (b) data block symbols, (c) auxiliary + data block symbols. The reported write energy is the average for 200 million random data blocks. The PCM memory line is 512-bits.
}\label{Six_four_random}\vspace{-.5em}
\end{figure*}
\begin{figure*}[!t]
\centering
\includegraphics[width=0.725\textheight]{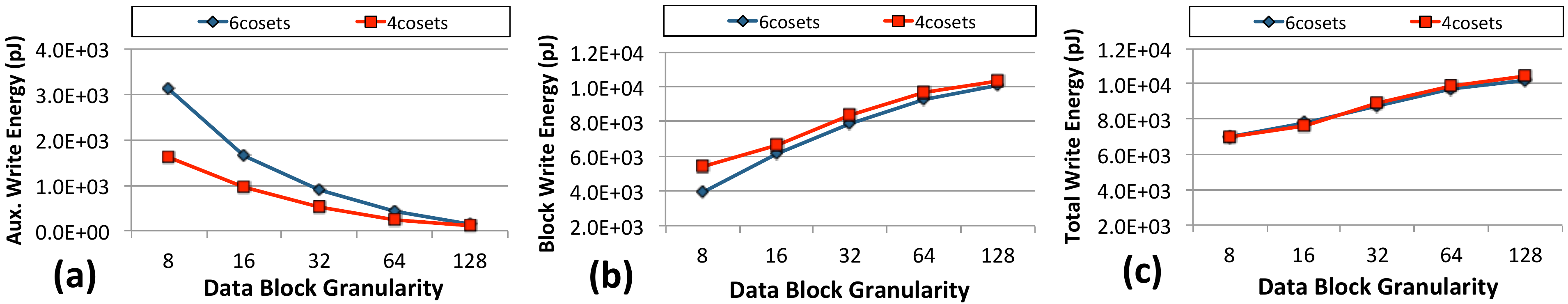}\vspace{-0.5em}
\caption{ Write energy analysis for (a) auxiliary symbols, (b) data block symbols, (c) auxiliary + data block symbols. The reported write energy is the average for SPEC2006 and PARSEC benchmarks. The PCM memory line is 512-bits.
}\label{Six_four_biased}\vspace{-1.25em}
\end{figure*}

The coset candidates used in~\cite{wang2011energy} to encode a memory line are based on mapping the two most frequent symbols in a memory line into the two low energy states while maintaining the original data block as much as possible. Assuming that any two of the four symbols can appear more frequently in any particular memory line, the encoding provides $C^{2}_{4}=6$ different mappings of symbols to states, which is equivalent to using six possible coset candidates in the encoding. Of course, 3-bits (two symbols) are needed for each memory line to record the particular candidate used in the encoding.

Note that the above logic used to select the six coset candidates is suitable for random data since it assumes that in any memory line, any two of the four symbols can appear more frequently in that line. However, it is well documented~\cite{ekman2005robust, balakrishnan2003exploiting,alameldeen2004frequent} that in real workloads, the two symbols `00' and `11' appear much more frequently than the other two symbols because many data words contain long runs of 0's or 1's. 
For example, zero is most commonly used to initialize data, to represent NULL pointers or false Boolean values, and to represent sparse matrices. On the other hand, long sequences of 1's appear in the representation of negative signed numbers. 
We will take advantage of this knowledge to propose four carefully selected coset candidates and compare the performance of this encoding, called `4cosets,' with that of the encoding
proposed in~\cite{wang2011energy}, called `6cosets.'
Note that by reducing the number of coset candidates from six to four, we reduce the auxiliary information needed to keep track of the coset candidate used from four bits (two symbols) to two bits (one symbol).
\renewcommand{\arraystretch}{1.25}
\begin{table}[!b]
\centering\vspace{-1em}
\caption{ Four coset candidates for mapping two bit patterns to the four energy states of a MLC PCM.}
\label{four_coset_candidates}\vspace{-1.25em}
\scalebox{0.9}{
\begin{tabular}{|l|l|l|l|l|l|}
\hline
\multirow{2}{*}{State} & \multirow{2}{*}{Write energy \cite{bedeschi2009bipolar}} & \multicolumn{4}{l|}{Coset candidate mapings of symbols to states} \\ \cline{3-6} 
& & Coset C1 & Coset C2 & Coset C3 & Coset C4 \\ \hline
S1 & \;\;\;\;\;\;\;36+0 \;\;\;pJ & \;\;\;00 & \;\;\;11 & \;\;\;11 & \;\;\;11 \\ \hline
S2 & \;\;\;\;\;\;\;36+20\ \;pJ & \;\;\;10 & \;\;\;00 & \;\;\;01 & \;\;\;00 \\ \hline
S3 & \;\;\;\;\;\;\;36+307 pJ & \;\;\;11 & \;\;\;10 & \;\;\;00 & \;\;\;01 \\ \hline
S4 & \;\;\;\;\;\;\;36+547 pJ & \;\;\;01 & \;\;\;01 & \;\;\;10 & \;\;\;10 \\ \hline
\end{tabular}
}\end{table}%
Table~\ref{four_coset_candidates} shows the symbol-to-state mapping for the four proposed coset candidates. The first candidate, C1, represents the default symbol-to-state mapping. Candidates C2 and C4 map `11' and `00' to the two states with the lowest write energy to take advantage of the fact that sequences of consecutive 0's and consecutive 1's are common in memory traces of real applications. Candidate C3 is chosen so that, when combined with C1, any of the four symbols will be mapped to the two states with the low write energy, either in C1 or in C3. This will be useful for random patterns that do not exhibit any bias.

To compare the effectiveness of the proposed 4cosets encoding with the 6cosets encoding proposed in \cite{wang2011energy}, we plot in
Figure~\ref{Six_four_random} the write energy for both encodings for 200 million random data blocks with granularity varying from 128-bits down to 8-bits.
Because it uses more candidates and has more options for reducing the write energy, 6cosets achieves write energy reduction in the data symbols more than 4cosets. The energy consumption for the auxiliary symbols is also lower for 6cosets than 4cosets despite the fact that 4cosets uses only one auxiliary symbol per data block while 6cosets uses two.
The reason is that for 6cosets, we use the six state combinations of the two auxiliary symbols that require the least write energy among the 16 possible state combinations of the two symbols. For 4cosets, all four states of the auxiliary symbol, including the two high write energy states, have to be used to identify the candidate used in the encoding.

The advantage of 6cosets vanishes when we compare the two schemes for real benchmarks as shown in Figure~\ref{Six_four_biased}.
The figure shows that 6cosets still has an advantage with respect to the write energy of data symbols. However, the energy to write the auxiliary symbols is lower in 4cosets than in 6cosets because it uses only one auxiliary symbol rather than two, and it uses the two low energy states of the auxiliary symbol to represent the most commonly used coset candidates, C1 and C2.
As a result, the total write energy in Figure~\ref{Six_four_biased} shows that the two sources of the write energy make a suitable trade-off such that the write energy of 4cosets is almost equal to that of 6cosets for a wide range of data block granularities.

\textbf{We conclude} that both 4cosets and 6cosets consume roughly the same write energy for real workloads. More importantly, 4cosets reduces the number of auxiliary symbols by 50\%, which is a large advantage when the memory line is to be compressed to make room for the auxiliary symbols.
\vspace{0.25em}
\section{\textbf{Word Level Compression~(WLC)}}
In the last section, we argued that by using 4cosets, rather than 6cosets, the number of auxiliary symbols is reduced by 50\%. 
Still, the main obstacle for using fine data block granularity (64-bits or smaller) is that it needs one auxiliary symbol per data block.
For example, at a 16-bit granularity, a 512-bit (256 symbol) memory line needs 64 auxiliary bits (32 auxiliary symbols), which is a 12.5\% overhead. For 8-bit and 32-bit granularities, the overhead is 25\% and 6.5\%, respectively. The question that we answer in this section is:\textit{
can we find a lightweight compression scheme that 1) successfully compresses a memory line with a high probability, 2) reduces the memory line size by 6.25\%, 12.5\% or 25\% to make room for the auxiliary symbols, and 3) does not disturb the biased bit patterns in a memory line to preserve the effectiveness of differential writes?}

\rami{To answer the above question, we apply different existing schemes for compressing 512-bit memory lines to different SPEC2006 and PARSEC workloads. The results show that FPC+BDI compression \cite{alameldeen2004frequent,pekhimenko2012base} only compresses 30\% of the memory lines, which does not satisfy the first constraint specified above. In contrast, the recently proposed Coverage-Oriented Compression (COC)~\cite{kim2015frugal} highly compresses cache lines by utilizing 28 different variable length compressors, which satisfies the first constraint. Unfortunately, COC does not satisfy the third constraint because its variable length encoding disturbs the biased bit patterns in a memory line. }
However, by inspecting each 64-bit memory word, we found that the $k$ most significant bits~(MSBs) of most words, for some $k$, have the same binary value and thus can be represented by only one bit. Following this new observation, we propose a simple and low-overhead Word Level Compression~(WLC) that compresses a 512-bit memory line as long as the $k$ MSBs of all its eight 64-bit words are compressible. 
\begin{figure}[!t]
\centering\vspace{0.05em}
\includegraphics[width=0.375\textheight]{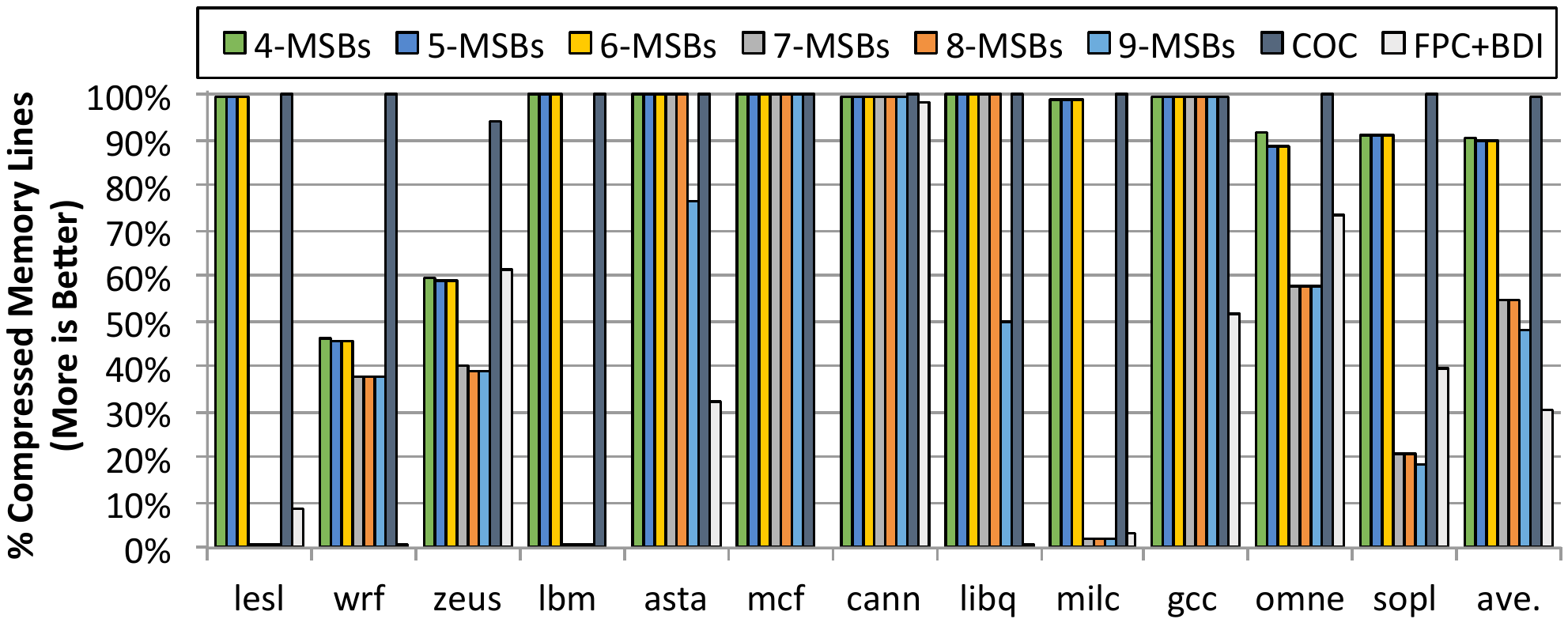}
\caption{ \rami{Comparison of the percentage of compressed memory lines by WLC, COC~\cite{kim2015frugal} and FPC+BDI~\cite{pekhimenko2012base}.%
}}\label{compression_comparison}\vspace{-1.25em}
\end{figure}
\begin{figure*}[!t]
\centering
\includegraphics[width=0.75\textheight]{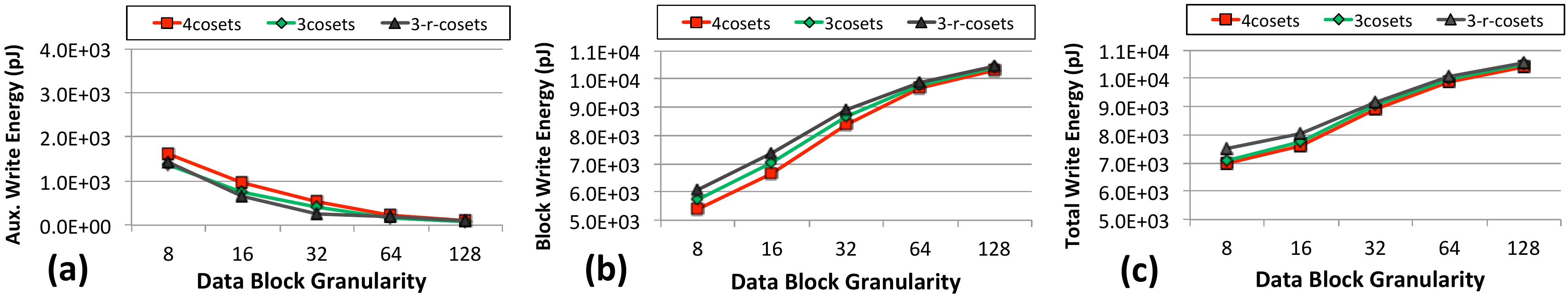}
\caption{ Write energy analysis for (a) auxiliary symbols; (b) data symbols, (c) auxiliary + data symbols. 
The average write energy are reported for the SPEC2006 and PARSEC benchmarks.
}\label{4_3_3_restrcited_no_compression}\vspace{-1.15em}
\end{figure*}


Figure~\ref{compression_comparison} shows the percentage of memory lines compressed by COC, BDI+FPC and WLC when $k$ is changed between 4 and 9. For 4MSBs, 5MSBs and 6MSBs, WLC compresses more than 91\% of the memory lines. With the 6 MSBs replaced by one bit in every 64-bit word, 5/64 = 7.8\% of a memory line is reclaimed and made available to store auxiliary symbols. Clearly, this is enough to store the auxiliary symbols for the 32-bit granularity, but not for smaller granularities.
Figure~\ref{compression_comparison} also shows that for $k$ > 6, WLC compresses 54\% of the memory lines. Although, with $k$ = 9, we can save \rami{8/64=12.5\%} of a memory line, which is enough to store the auxiliary symbols for 16-bit granularity, only 48\% of the memory lines can be compressed and encoded at this value of $k$ while the remaining lines will be written without any encoding. This motivates the idea of restricted coset encoding to decrease further the
amount of auxiliary information.

\section{\textbf{Restricted Coset Coding}} 
In traditional coset encoding, a coset candidate is selected independently for each data block to minimize the write energy for that block. 
In this section, we introduce a restricted coset encoding which mandates a correlation between the use of coset candidates in a number of consecutive data blocks. This restriction reduces the auxiliary information and does not largely affect the energy minimization capability because the bit patterns of consecutive words are usually similar.

We illustrate the concept of restricted coset encoding by a simple example. Assume that we only use the first three coset candidates, C1, C2 and C3, discussed in Section~\ref{coset}. Instead of allowing the flexibility of using C1, C2 or C3 independently in each data block, we can group the cosets into two groups, `C1,C2' and `C1,C3', and force any data block in a memory line to either use one of C1 and C2 in the encoding or to use one of C1 and C3. 
For example for a 16-bit encoding granularity, there are 32 data blocks in a 512-bit memory line. Restricted coset encoding proceeds as follows: 1) use the two candidates C1 and C2 to encode each of the 32 data blocks, 2) use the two candidates C1 and C3 to encode each of the 32 data blocks and 3) select the better of the encodings produced in steps 1 and 2. 
Of course, this restricted method needs one global auxiliary bit per memory line to determine the coset group used in that line, in addition to one auxiliary bit per data block, for a total of 33 auxiliary bits (17 symbols) per memory line. This is fewer than the unrestricted encoding which needs 64 auxiliary bits (32 symbols) per memory line, two bits per data block.

To explain the ramification of restricting the use of cosets, we recall from the previous section the justifications for choosing the three candidates C1, C2 and C3. C2 is useful for biased data with many sequences of consecutive 0's or 1's, while C3 is useful for non-biased data. Because of data locality, we can expect consecutive words in the memory line to either be all biased or not. In the former case, not using C3 will not hurt much, and in the latter case, not using C2 will not hurt much.

To evaluate the effect of restricting the use of cosets on the write energy, we plot in Figure \ref{4_3_3_restrcited_no_compression} the write energy of 4cosets, 3cosets (that unrestrictedly uses candidates C1, C2 and C3) and the restricted coset coding (called 3-r-cosets). We draw two observations from this figure. First, 3cosets only slightly increases the write energy over 4cosets. Second, reducing the number of auxiliary information by the proposed restricted method increases very little the write energy relative to 4cosets. The main advantage of restricting the coset candidates will be clear in the next section where we use WLC to make room in the memory line for embedding the auxiliary information.

\section{\textbf{WLCRC: Integrating WLC with Restricted Coset Encoding}}

In this section, we will use WLC to make enough room in the memory line to store the auxiliary symbols of the restricted coset encoding. Because of the reduction in the auxiliary information needed for 3-r-cosets, WLC will be able to provide the necessary room in 91\% of 
memory lines to embed the auxiliary symbols. A global bit (symbol) per memory line will be used to flag the lines that cannot be compressed. Those lines will be written in memory without encoding.

Figure~\ref{diagram}(a) shows the WLC that compresses the 6 most significant bits of each 64-bit word of a 512-bit memory line. In this figure, each row represents the 64-bits, $b^{i}_{63}, ..., b^{i}_{0}$, of each of the eight words, $w^{i}$, $i=0,...,7$. WLC compresses the memory line as long as all six MSBs, $b_{63}, ... , b_{58}$, of \textbf{each word} are `000000' or `111111'. Thus, it splits each word into two parts: the five reclaimed bits, $b_{63}, ..., b_{59}$, and the data bits $b_{58}, ... , b_0$, which are not changed by the compression. When decompressing the word, bit $b_{58}$ is extended to the reclaimed bits, similar to sign extension. The five reclaimed bits will be used to store the auxiliary bits of the 3-r-coset encoding at 16-bit granularity.

Figure~\ref{diagram}(b) shows the format of the restricted coset encoding at a 16-bit granularity. Specifically, each 64-bit word is divided into 4 data blocks, $`b_{58}, .., .b_{48}$', $`b_{47}, ..., b_{32}$', $`b_{31}, ..., b_{16}$', and $`b_{15}, ..., b_{0}$'.
To record the coset restriction used for encoding each 16-bit data block in a word, we use bit $b_{63}$ to determine which group of $coset_{C1,C2}$, or $coset_{C1,C3}$, is used to encode the four 16-bits data blocks in the 64-bit word.
Then, the four bits, $b_{62}, ... , b_{59}$, are used to identify which coset candidate (restricted by the specified group) is used in each data block. 
\begin{figure}[!t]
\centering
\includegraphics[width=0.375\textheight]{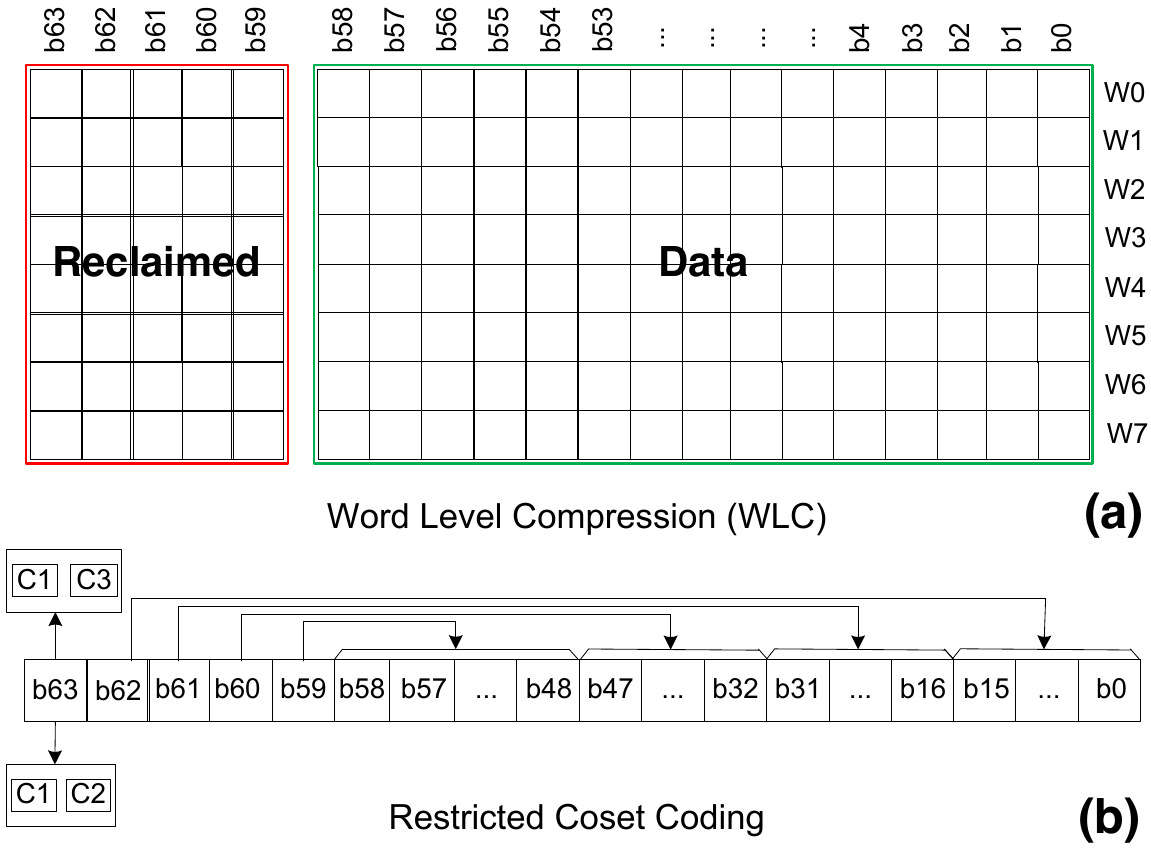}\vspace{-0.25em}
\caption{ (a) Word Level Compression (WLC), (b) Restricted coset coding at 16-bit granularity.
\vspace{-2.5em}
}\label{diagram}
\end{figure}
\rami{Algorithm~\ref{Encoder} is a pseudo-code for WLCRC-16 where the eight 64-bit words, $w^{i}$, $i=0,...,7$, are independently encoded in parallel when the memory line is compressible~(Line~1-2). To encode a word, $w^i$, it is divided into 4 sub-words $w^{i}_{j}$, $j=0,...,3$ (Line~3), and the sub-words are encoded in parallel using the three cosets C1, C2 and C3. Then, the energy cost, $cost_k(w^i_j)$, of encoding $w^i_j$ using C$k$ is computed for $j=0,...,3$ and $k=1,2,3$. This allows the estimation of $cost_{1,2} (w^i)$ and $cost_{1,3} (w^i)$, which are the costs of encoding $w^i$ using C1 or C2 and using C1 or C3, respectively (line 4). Finally, the encoding with minimum cost is selected to be written in memory.}

Note that, driven by the compressed format, we applied the coset restriction to the data blocks in a 64-bit word, rather than to the entire memory line, as described in the previous section. Hence, our proposed encoding, called WLCRC, applies only to data blocks at granularities of 8, 16, 32 and 64 bits. However, to apply WLCRC at 8-bit granularity, eight bits must be reclaimed by WLC from each word. To apply it at 32-bit granularity, only three bits must be reclaimed. At 64-bit granularity, WLCRC is identical to the unrestricted 3cosets encoding in which also two bits need to be reclaimed.

Finally, we note that WLC can be integrated with unrestricted 3cosets or 4cosets encodings, as long as WLC can reclaim enough bits to embed the auxiliary bits for the encoding. For example, to use WLC with 4cosets at data block granularities of 8, 16, 32 or 64 bits, WLC has to reclaim 16, 8, 4 and 2 bits per word, respectively. Note, however, that according to Figure~\ref{compression_comparison}, as long as the number of reclaimed bits per word is less than 6 ($k$-MSBs with $k < 7$), WLC compresses 91\% of the memory lines. Otherwise it compresses fewer than 55\% of the lines. In summary, \textit{the selection of data block granularity and restricted/unrestricted encoding is a trade-off between the encoding overhead and the write energy reduction.}\vspace{0.25em}

\subsection{\textbf{WLCRC Architecture}}

\setlength{\textfloatsep}{0pt}
\begin{algorithm}[t]
\Begin{
\For{$w^{i}, i=0,...,7$, in Parallel,}{
Divide $w^{i}$ into four sub-words $w^{i}_{j}$, $j=0,...,3$\\
Encode $w^{i}_{j}$ using C1,C2 and C3 in parallel
$cost_{1,2}(w^i)=\sum_{j=0}^{3} min \{cost_{1}(w^i_j), cost_{2}(w^i_j)\}$
$cost_{1,3}(w^i)=\sum_{j=0}^{3} min \{cost_{1}(w^i_j), cost_{3}(w^i_j)\}$
\\If $(cost_{1,2}(w^i) < cost_{1,3}(w^i))$ encode $w^i$ using C1/C2 else encode $w^i$ using C1/C3.
}
}
\caption{Pseudo-code for WLCRC-16 applied to a compressible memory line.}\label{Encoder}
\end{algorithm}
Figure~\ref{architecture} shows the on-chip architecture of WLCRC compression+encoding and decoding+decompression for a data block granularity of 16. The 512-bit line from the memory controller is sent to the WLC module to check whether it is compressible or not. 
If the line is compressible, WLC enables the encoder to activate eight restricted coset encoding modules. When differential write is used, each compressed 64-bit word out of WLC is differentiated with the corresponding 64-bits from the currently stored memory line and the difference is used in a restricted coset module to compute the encoded word to be written into memory.
If WLC cannot compress the data line, the uncompressed, unencoded line is compared with the memory current data and the difference is written to memory. One auxiliary bit is used to differentiate encoded from non-encoded lines, which means that an additional symbol must be stored with the 256 symbols of the memory line. Consequently, the total encoding space overhead is~<~0.4\%. Note that the leading 6cosets scheme stores two auxiliary symbols with each memory line, which is double the space overhead of WLCRC.
\begin{figure}[!t]
\centering
\includegraphics[width=0.35\textheight]{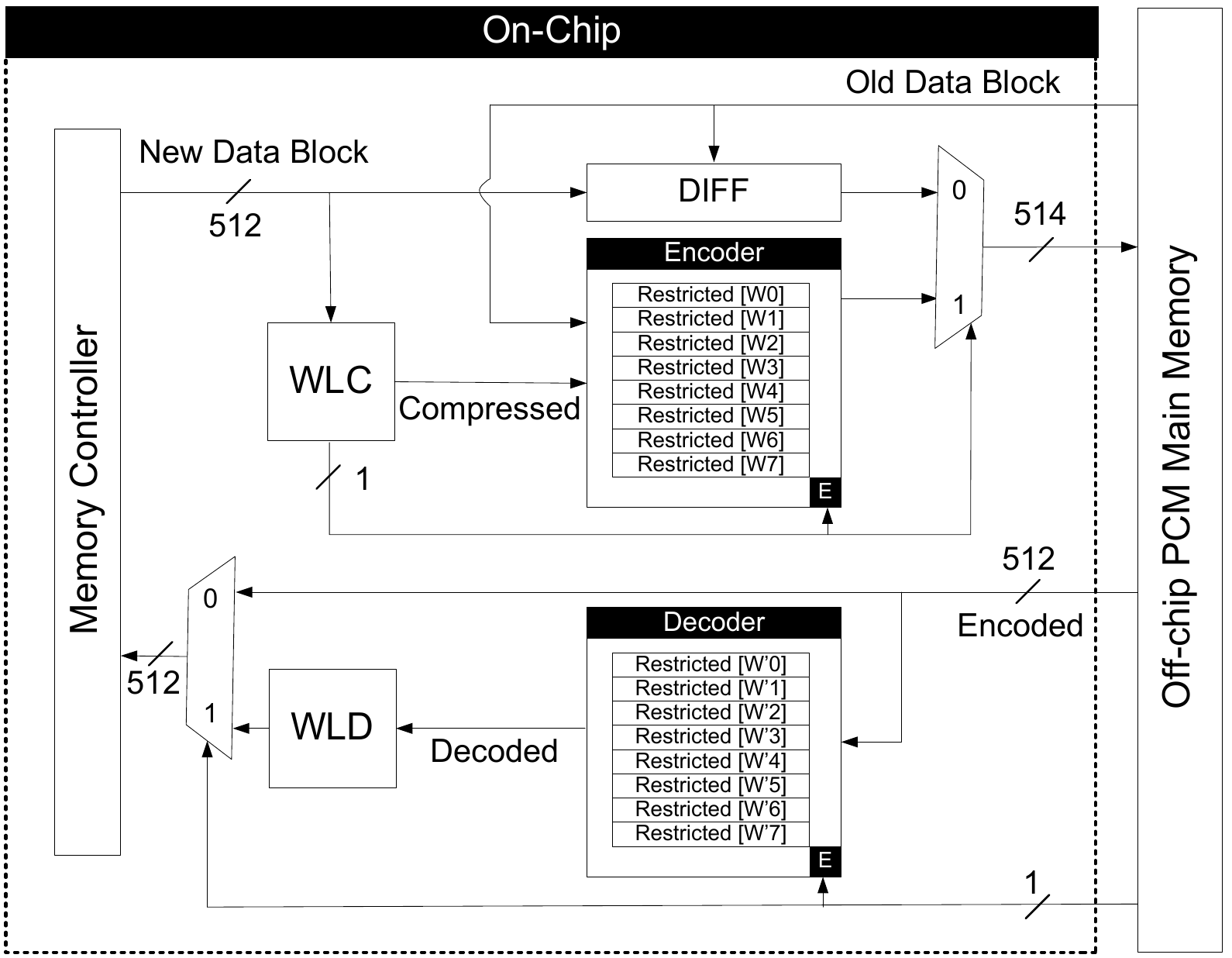}
\caption{ On-chip WLCRC architecture for 16-bit granularity.
\vspace{0.15em}
}\label{architecture}
\end{figure}

The eight 64-bit encoders operate in parallel. Each encoder splits the word into 16-bit blocks and for each block, the writing cost is estimated when each of the candidates, C1, C2 or C3 is used for mapping the symbols to the cell's states.
Note that to encode the four data blocks in parallel, the most significant block, $b_{58}, ... , b_{48}$, contains 11 bits rather than 16 bits since bits $b_{63}, ..., b_{59}$ are not known before the encoding. It is possible, however, to consider all 16 bits, $b_{63}, ... , b_{48}$, in the encoding process if we encode the most significant block after the encoding of the other three blocks is completed, which will increase the encoding (and similarly the decoding) delay. We chose the fully parallel solution.

The decoding follows the reverse structure of the encoding. Specifically, it first checks whether the memory line has been compressed/encoded or not. If yes, the decoder decodes the eight words and then a WLD module decompresses the decoded words. 
The decoding process is simple as the most significant bit of each 64-bit word, $b_{63}$, determines the coset group that had optimized that word in the encoding process. Then, the four bits, $b_{62}, ..., b_{59}$ determine the coset candidate that should be used to decode the corresponding 16-bit block.

\subsection{\textbf{Hardware Overhead}}\label{hardware}
In this section, we evaluate the delay, power, area, and energy of WLCRC-16. Verilog implementations were synthesized using Synopsys Design Compiler targeting a 45nm FreePDK standard cell library~\cite{FreePDK45}. The WLCRC implementation assumes 512-bit memory lines requiring eight encoding modules to simultaneously encode the compressed words by WLC. \rami{We assume that the additional encoding bits added to the 512-bit memory line are handled through a wider main memory interface.} Our results show that the WLCRC modules incur an area overhead of 0.0498${mm^{2}}$, which is negligible compared to the PCM main memory area. Moreover, the delay of WLCRC modules is 2.63${ns}$ and 0.89${ns}$ during a write and read, respectively. The energy consumption of the WLCRC modules is 0.94${pJ}$ and 0.27${pJ}$, per write and read memory line access, which is negligible compared to the write energy consumed by cell programming. Note that the WLC compression/decompression portion of the design is very small compared to the encoding/decoding portion, requiring only 0.0002$mm^{2}$ area, 0.13${ns}$ delay, and 0.0017${pJ}$ of energy. 

\section{\textbf{Experimental Settings}}\label{sec:setting}
\subsection{\textbf{Experimental Configuration}}\label{sec:configuration}
To conduct experiments, we developed a trace driven simulator. The input traces to our simulator were collected with Virtutech Simics~\citep{magnusson2002simics}. As it is widely assumed that PCM employs differential write, or writing bits only when the value differs from the previously stored value, for each memory write transaction the traces store both the value to be stored as well as the value to be overwritten. 

For trace generation, our simulations assume an 8-core 4GHz chip multiprocessor. 
Each core has a 2MB private L2 cache. We model a 32GB PCM main memory with two channels; each channel has two DIMMs and each DIMM has eight chips and 16 banks. In general, the read queue is given a higher priority than the write queue. However, to avoid starvation, when the write queue exceeds 80\% of capacity, writes are serviced ahead of reads. For write energy evaluation, we scaled the write energy reported from an MLC PCM prototype at the 90nm process node~\cite{bedeschi2009bipolar,wang2011energy}. 
All studied schemes are implemented on top of differential write~\cite{zhou2009durable}. We used a `single' RESET and multiple SET iteration-based programming strategy~\cite{braga2010voltage} to increase programming accuracy in our evaluation\footnote{An alternative programming scheme is to use one SET pulse and multiple RESET pulses~\cite{joshi2011mercury}. Because of reliability concerns such as resistance drift and difficulty in controlling the melting process, we selected the `one SET - multiple RESETs' scheme.}. If the cell value does change and requires a write, it consumes the RESET energy of about 36$pJ$. Then depending on the cell value, SET operations may ensue to change its resistance requiring between 20$pJ$ and 547$pJ$. 

The write disturbance error rates~(DER) of MLC PCM states when the adjacent cell is being written (modeled by the RESET operation) are also extracted from the literature~\citep{jiang2014mitigating}. Thus, an idle cell in the minimum resistance state is assumed to be error free as the high heat of the RESET process will not increase its resistance. Note that the lowest energy states, S1 and S2, are the highest and lowest resistance states, respectively. RESET places the cell in the highest resistance state (S1) and a short high write current can place the cell in the lowest resistance state (S2) (immune to write disturbance), similar to SLC PCM. States S3 and S4 require many more precise SET operations to achieve a resistance between the high and low energy state, making them require high write energy as well as making them susceptible to write disturbance when idle. All schemes are compatible with the standard `Verify-n-Restore' approach~\cite{dong2011adams} to correct disturbance errors that may have occurred. Detailed simulation parameters are recorded in Table~\ref{tab:configuration}.

To evaluate endurance, we counts the average number of updated cells per write request since fewer RESET operations leads to higher cell endurance. To evaluate write disturbance, we count the number of idle cells disturbed by neighboring cells that need to be updated in the write request. The write disturbance phenomenon happens during the RESET process that generates high heat and can potentially disturb adjacent cells in states S1, S3 and S4 with the probabilities shown in Table~\ref{tab:configuration} based on a 20nm technology node~\cite{jiang2014mitigating}. 

\setlength{\textfloatsep}{10pt plus 1.0pt minus 2.0pt}
\renewcommand{\arraystretch}{1.45}
\begin{table}[!t]
\centering
\caption{System configuration
}\vspace{-1em}
\label{tab:configuration}
\scalebox{0.95}{
\begin{tabular}{|l|l|}
\hline
CPU & \begin{tabular}[c]{@{}l@{}}8-core, 4GHz, single-issue \\
\end{tabular} \\ \hline
L2 Cache & \begin{tabular}[c]{@{}l@{}}private 2MB, 8-way\\ 64B line, write-back\end{tabular} \\ \hline
\begin{tabular}[c]{@{}l@{}}32GB\\ MLC PCM\\Main Memory\end{tabular} & \begin{tabular}[c]{@{}l@{}}2 channels\\ 2 DIMMs per channel\\ 16 banks per DIMM, 32-entry, \\64B line write pausing scheduling\end{tabular} \\ \hline\hline
\begin{tabular}[c]{@{}l@{}}MLC PCM \\
36pJ RESET\\
Energy\end{tabular}
& \setlength{\tabcolsep}{1pt}
\begin{tabular}[c]{@{}l|l@{}} 
Set Energy~\cite{wang2011energy,bedeschi2009bipolar} & Disturbance Rate~\cite{jiang2014mitigating}\\\hline
S1:~0pJ&~DER:~12.3\%\\
S2:~20pJ&~DER:~0.0\%\\
S3:~307pJ&~DER:~27.6\% \\
S4:~547pJ&~DER:~15.2\%\end{tabular}\setlength{\tabcolsep}{7pt}
\\ \hline
\end{tabular}
}
\end{table}
\subsection{\textbf{Workloads}}\label{sec:workload}
In order to study the impact of our scheme on write energy of MLC PCM, we selected memory intensive workloads. In particular, we
include twelve write-intensive benchmarks from SPEC CPU2006 and supplement them with \textit{canneal} from PARSEC. We selected only the canneal workload from PARSEC because most PARSEC benchmarks are computation intensive and in most cases also have a very small memory footprint. To be consistent with the SPEC CPU workloads, canneal was executed in our experiments in single-threaded mode and with the largest `native' data input that resulted in a 940MB memory footprint. For SPEC CPU2006, we use the large `reference' inputs that are designed to stress the system.\vspace{-0.25em}
%
\begin{figure*}[!b]
\centering\vspace{-0.8em}
\includegraphics[width=0.685\textheight]{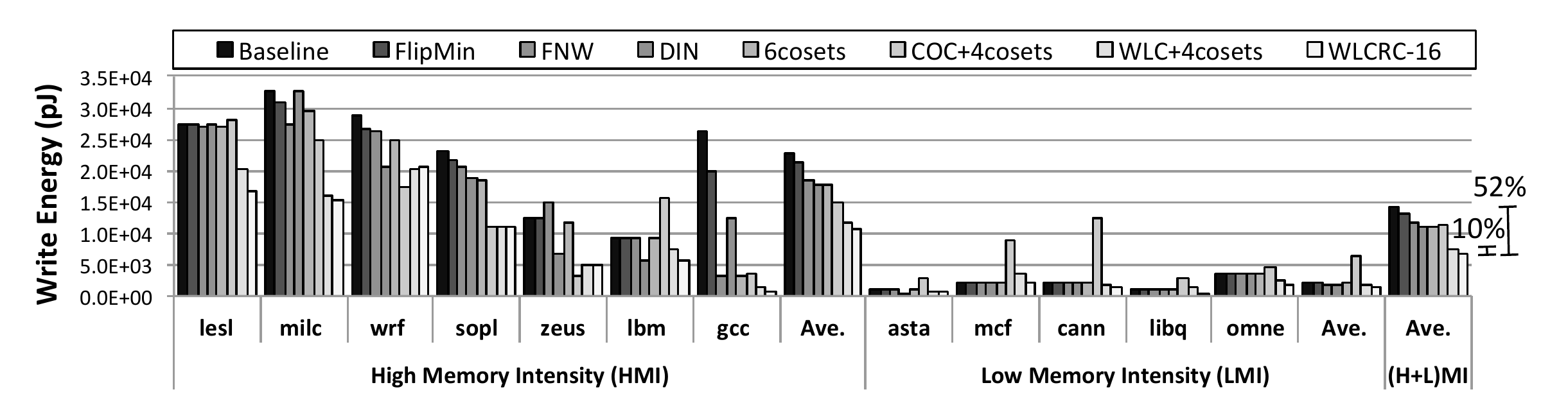}\vspace{-0.7em}
\caption{ \rami{Comparison of write energy for various schemes on SPEC CPU2006 and PARSEC inputs.}}\label{workloads_energy}
\end{figure*}
\section{\textbf{Evaluation}}\label{sec:evaluation}
To evaluate the effectiveness of WLC and the restricted coset coding, we compared the following schemes:\vspace{0.25em}
\begin{description}
\item [Baseline:]~~~This scheme just uses standard differential write for energy reduction when writing a 512-bit memory line into MLC PCM. 
\vspace{0.25em}\item [FlipMin]~~\;\cite{jacobvitz2013coset}\textbf{:}~~This scheme uses two symbols (four auxiliary bits) per memory line for 16 coset candidates, generated using the technique in~\cite{seyedzadeh2016improving}, operating on a 512-bit memory line. Note that this scheme, as well as the next scheme, FNW, were proposed for SLC PCM and were adapted in our implementation for MLC PCM.
\vspace{0.25em}\item [FNW]\cite{cho2009flip}\textbf{:} This scheme
selects the original data block or its complement, depending on which one uses less write energy. A single auxiliary bit is enough to indicate that a data block is complemented. Thus, to match the space overhead of FlipMin which uses two symbols (four auxiliary bits) per 512-bits memory line, we partition the memory line into 128-bit blocks that can be inverted independently with FNW.
\vspace{0.25em}\item [DIN]\cite{jiang2014mitigating}\textbf{:} This scheme uses a 3-to-4-bit code word mapping to remove high energy states. Write disturbance errors are mitigated by a 20-bit BCH code to correct two write errors in the write verification process. To avoid the space overhead of this encoding, it is applied only to 512-bit memory lines that can be compressed with FPC+BDI to at most 369 bits. 
DIN was originally proposed to reduce write disturbance.
\vspace{0.25em}\item [6cosets]~\cite{wang2011energy}\textbf{:}~This scheme uses six coset candidates to map any two of the four symbols to the low energy states S1 and S2. Thus, it also incurs a space overhead of two auxiliary symbols (four bits) per 512-bit memory line.
%
%
\vspace{0.25em}\item [COC~\cite{kim2015frugal}]~~~+4cosets:~This scheme uses COC along with directly applying the four coset candidates shown in Table~\ref{four_coset_candidates}. The encoding is applied at 16-bit or 32-bit granularity for lines that are compressed to at most 448 bits or 480 bits, respectively.
\vspace{0.25em}\item [WLC+4cosets:]~~~~~~~~\color{white}a\color{black}This scheme uses WLC along with directly applying the four coset candidates shown in Table~\ref{four_coset_candidates}. It requires a space overhead of one symbol per memory line to indicate if the memory line is compressible or not. Unless stated otherwise (in Section~\ref{sens_analysis}), the default WLC+4cosets encoding granularity is 32-bit blocks. 
\vspace{0.25em}\item [WLCRC:]~~~This scheme, WLC with the restricted coset encoding, uses the first three coset candidates shown in Table~\ref{four_coset_candidates}.
The default WLCRC granularity is for 16-bit blocks, denoted as WLCRC-16.  \vspace{0.25em}

\end{description}

Note that for \rami{COC+4cosets}, WLC+4cosets and WLCRC-16 encoding techniques, when \rami{COC} and WLC cannot sufficiently compress the block, the original, uncompressed 512-bit memory line is written. Because the auxiliary symbol must only record the compression state, even though it can store four states, we select only the two lowest energy states for this purpose. Moreover, since \rami{COC} and WLC compress more than 90\% of memory lines, we flagged the `compressed' state with the lowest energy state. 

In the following sections, we compare these enumerated schemes for their write energy, their average number of updated cells per write request, and their average number of write disturbance errors per write request for as close to an ISO-overhead comparison as possible. 
In general, these schemes are categorized into two groups. The first group, including FlipMin,~FNW, and 6cosets, augments the encoding space for an entire memory line 
to reduce the energy, while the second group, including DIN, \rami{COC+4cosets}, WLC+4cosets, and WLCRC-16, use compression techniques in order to allow encoding at a finer block granularity to reduce write energy. \vspace{0.25em}

\subsection{\textbf{Write Energy}}\label{sec:write_energy_wrokload}
Figure~\ref{workloads_energy} compares the write energy for different schemes. While FNW is superior to FlipMin, in part due to its ability to operate on a smaller block size, 6cosets performs the best among the schemes designed to operate on the full memory line. Interestingly, on average DIN, which operates on the smallest block size, performs close to 6cosets, but its effectiveness is much more benchmark dependent. This is likely due to the varied effectiveness of the FPC+BDI compression that enables DIN encoding within each particular workload. 

In contrast, word level compression is extremely effective and consistent in reducing energy. In particular, WLC+4cosets provides a 46\% improvement over the baseline and 32\% improvement over the leading 6cosets approach. Further decreasing the block granularity at the expense of the coset flexibility provides a significant additional improvement. WLCRC-16 reduces the write energy by 10\% over WLC+4cosets and increases the improvement over the baseline to 52\% while providing an overall improvement of 39\%, 39\%, and 48\% versus 6cosets, DIN, and FlipMin, respectively. For all workloads, including non-intensive memory applications, WLCRC-16 reduces the write energy on average versus other schemes. Moreover, Figure~\ref{workloads_energy} shows that, as expected, write energy grows considerably for intensive workloads, such as \textit{milc}, \textit{lesl}, and \textit{sopl,} while the effectiveness of WLC and, in particular, WLCRC-16 scales very well. For the high energy benchmark \textit{wrf} where 6cosets is not effective but DIN is effective, WLC-based schemes are still the best approach.

\begin{figure*}[!b]
\centering\vspace{-0.35em}
\includegraphics[width=0.685\textheight]{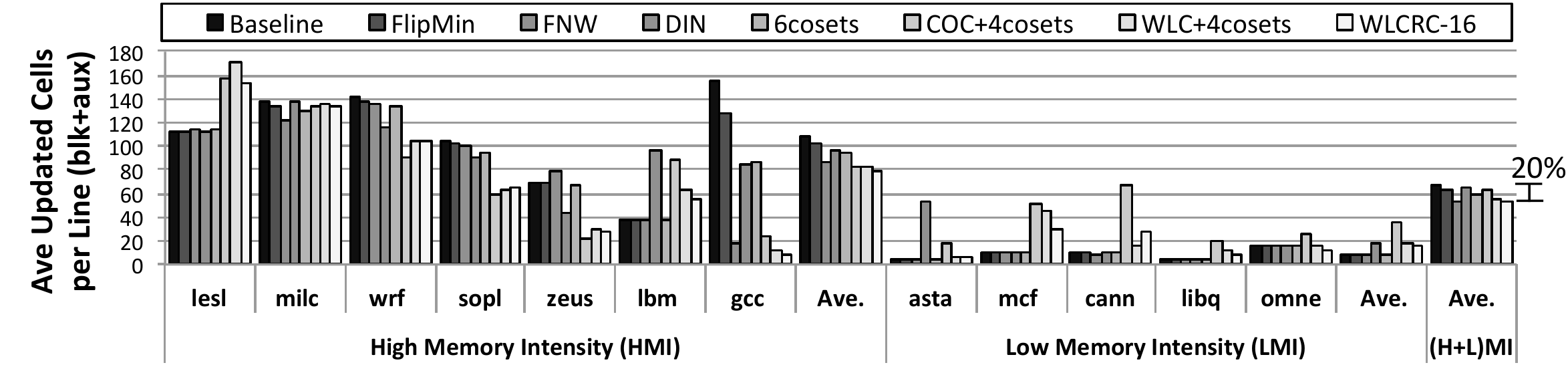}\vspace{-0.65em}
\vspace{-.05in}
\caption{ \rami{Average number of updated cells per memory line for SPEC CPU2006 and PARSEC inputs.}\vspace{-1.5em}}\label{workloads_endurance}
\end{figure*}
The effectiveness of the proposed techniques comes from several factors. First, they employ coset candidates that best map commonly occurring bit sequences to low energy states for different types of workloads. Second, WLC+4cosets and WLCRC achieve a small data block granularity for encoding, which can more precisely select the best coset candidates to map symbol encoding. Third, WLC compression can be applied to more than 91\% of the memory lines in these representative workloads, making coset encoding possible in a very high percentage of blocks. Fourth, contrary to compression techniques that significantly change the content of compressed data blocks even for relatively small changes in actual data, WLC only compresses a small fraction of the 64-bit word to create room for the coset auxiliary bits, retaining much of the temporal locality that makes differential writes effective. 
In contrast 6cosets and FlipMin operate at a large data block granularity~(512-bits) since they require a substantial increase in auxiliary information to operate at a granularity similar to WLC-based encoding. The additional auxiliary bits tend to work against the energy saved in the data block due to the random nature of the encoding. For example, decreasing the granularity for 6cosets from 512-bits to 16-bits increases the write energy ratio of the auxiliary symbols to the data symbols from 0.78\% to 12.5\%. The restricted coset method further decreases the number of auxiliary symbols, making encoding improvements to the data block more impactful. 

In contrast to DIN, which requires 25\% compression of the memory line to accommodate 3-bit to 4-bit expansion, restricted coset encoding requires only 7.8\% compression. Figure~\ref{compression_comparison} shows that more than 70\% of memory lines cannot be compressed for DIN while 91\% of memory lines are compressible with WLC. Moreover, the compression and BCH encoding employed by DIN increase symbol flips in the memory line, limiting the possible energy savings. 

The 10\% write energy reduction of WLCRC-16 versus WLC+4cosets is primarily due to the latter's need to operate on 32-bit blocks. For WLC+4cosets to operate at 16-bit granularity would require WLC to reclaim eight bits every 64-bit words rather than five bits for WLCRC-16. Unfortunately, the number of compressible memory lines reduces from 91\% to 48\% when eight rather than five bits are to be reclaimed by WLC, making WLCRC-16 much more effective. 

\rami{While COC+4cosets is somewhat effective in reducing the write energy for high memory intensity workloads, it tends to increase the write energy for low memory intensity workloads. 
Our analysis of the COC-4cosets encoded memory lines shows that it uses 16-bit data block granularity for most write requests. However, since PCM uses differential write to update only modified bits, it is important to ensure that compressors not increase the bit entropy of consecutive write requests. Because COC was not designed to preserve bit entropy, it often switches between the 28 different compressors, changing the data bit patterns from the original. In contrast, WLCRC-16 does not change the bits in the data except in only a few locations, which allows the differential write to take advantage of data locality. This is why, on average, WLCRC-16 uses 39\% less energy than COC+4cosets.

In summary, novelty of the WLC is that it is a simple compression mechanism that can, with high probability, compress memory lines enough to make room for auxiliary encoding bits, while preserving the bit location/locality of most of the bits. This is a crucial property for effective differential writes.
}
\begin{figure*}[!t]
\vspace{-.05in}
\centering
\includegraphics[width=0.675\textheight]{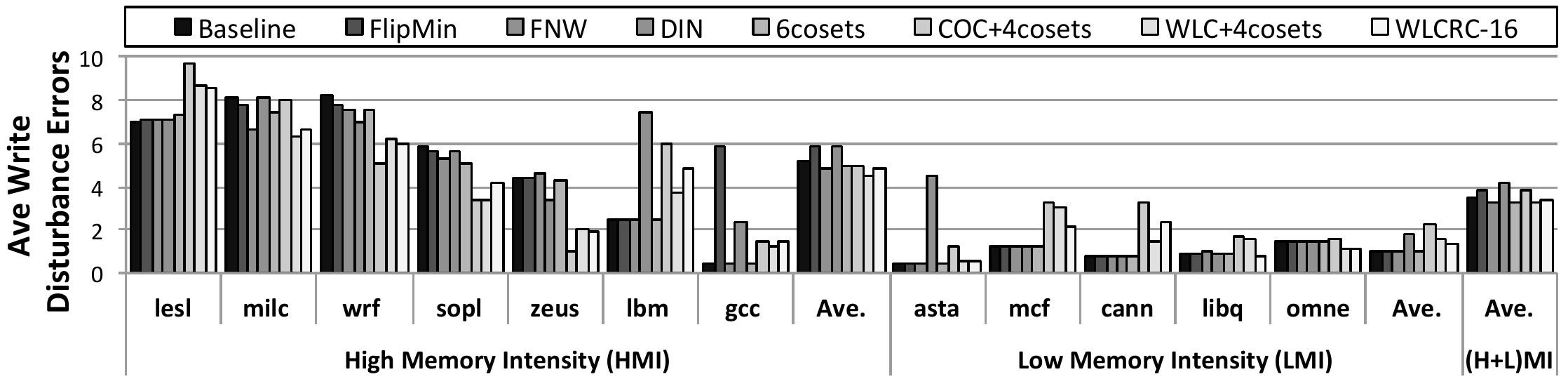}
\vspace{-.095in}
\caption{ \rami{Average number of write disturbance errors per memory line for SPEC CPU2006 and PARSEC inputs.}\vspace{-1em}
}\label{workloads_wd}
\end{figure*}
\subsection{\textbf{Endurance}}\label{sec:endurance_workload}
PCM main memory employs differential write to decrease the number of written cells primarily to save energy. However, the reduced numbers of writes also benefits endurance. Reducing the number of cells that are changed through intelligent encoding, such as WLC+4cosets and WLCRC-16, can further improve endurance. \rami{Figure~\ref{workloads_endurance} shows the average number of updated cells per write request. It shows that WLCRC-16 reduces the number of updated cells by 20\%, 17\%, 16\% and 11\% versus the baseline, FlipMin, COC+4cosets and 6cosets schemes, respectively.} However, the improvement or degradation in endurance varies highly for different benchmarks. For some workloads, such as \textit{wrf}, \textit{zeus}, \textit{gcc}, and \textit{sopl}, WLCRC-16 not only reduces write energy but also reduces the average number of updated cells, thus improving endurance. 
For other workloads such as \textit{lesl}, \textit{lbm}, \textit{mcf}, and \textit{cann}, WLCRC-16 more frequently maps high energy states to low energy states to reduce write energy but causes an increase in the number of updated cells compared to other schemes, thus harming endurance. 
Therefore, \rami{WLCRC-16 makes a trade-off between write energy and the number of updated cells for this group of workloads. However, on average WLCRC-16's endurance is considerably better than 6cosets, COC+4cosets and DIN and is on par with FNW.}
%
\subsection{\textbf{Write Disturbance}}\label{sec:write_disturbance_workload}
Write disturbance errors occur during the RESET process. The high heat of RESET (melting the material) can change the resistance of nearby idle cells that are not part of the actual write request. Write disturbance is unidirectional, so it can only decrease the resistance of other cells. Cells with the minimum resistance (S2) are thus immune to write disturbance. However, any RESET operation adjacent to a cell in states S1, S3, or S4 may still incur write disturbance.

Our results shown in Figure~\ref{workloads_wd} indicate that all schemes on average face write disturbance errors ranging from three to four every request to write a 512-bit memory line. For more memory intensive workloads such as \textit{lesl} and \textit{milc}, the average number of write disturbance errors across all schemes ranges between seven and nine. DIN compressed data blocks increase the number of cells written which increases write disturbance to be the highest among all the approaches. However, its 20-bit BCH code offsets this somewhat by correcting two disturbance errors. WLC+4cosets and WLCRC perform generally well, averaging around the minimum point for all benchmarks.

Part of the trends observed in Figure~\ref{workloads_wd} is the correlation between disturbance faults and the number of updated cells per write operation. When more cells are written, the likelihood of disturbing adjacent idle cells increases.

\rami{Since PCM uses differential writes, a memory line is always read before it is written. This allows for the detection of write errors by a "read-after-write" process, thus avoiding silent data corruption (SDC) due to write disturbance. It also allows for an iterative verify-and-restore (VnR) process~\cite{dong2011adams} which iterates until data is correctly written, thus eliminating detected uncorrectable errors (DUE). Consequently, any available Chipkill capability can be used for non-write-disturbance errors. It was shown in~\cite{jiang2014mitigating} that write disturbance errors can be completely removed if 3-5 iterations of VnR are used. Moreover, only the cells that are neighbors of the written cells are involved in each VnR iteration, which limits the effect on memory bandwidth and avoids resource starvation.
As indicated in \cite{jiang2014mitigating}, minimizing the probability of write disturbance (which WLCRC does) will improve performance because of the reduction in the number of VnR iterations. Finally, we note that although the different schemes differ in the average number of disturbances per line, the maximum number of disturbances per line changes very little across schemes (ranges between 15 and 17).} 

In summary, WLCRC improves write energy while achieving comparable endurance and write disturbance compared to the schemes specifically designed to improve these metrics.
%
%
%
\subsection{\textbf{Multi-objective Optimization}}
\rami{ The results in Figures~\ref{workloads_energy} and \ref{workloads_endurance} show that, for some applications with unbiased patterns, such as \textit{lesl} and \textit{lbm}, minimizing the write energy may increase the number of updated cells and result in degraded endurance. The main reason is that sometimes the coset candidate that minimizes energy actually increases the number of cells written into low energy states to avoid a relatively smaller number of writes into high energy states. It is possible, however, to select the encoding cosets based on a function that combines energy and endurance, thus sacrificing some energy improvement to attain better endurance. For example, recalling line 5 of Algorithm~\ref{Encoder}, if the difference between $cost_{1,2}(w^i)$ and $cost_{1,3}(w^i)$ is smaller than a threshold, $T$, then the encoding choice can be made based on the number of written symbols rather than energy.

We applied this multi-objective scheme to WLCRC-16 and successfully improved the endurance with a negligible sacrifice in energy saving. 
For example, when WLCRC-16 with $T$=1\% is applied to \textit{lesl} and \textit{lbm}, the average number of updated cells is reduced from 153 to 133 and from 55 to 49, respectively, while the write energy increased by less than 1\%. When we applied WLCRC-16 with T=1\% to all benchmarks, the number of updated cells decreased by 19\% (52 to 42) on average, while increasing the write energy from 6777pJ to 6885pJ. Relative to the baseline, applying the multi-objective optimization to WLCRC-16 increases the endurance improvement from 20\% to 35\% while resulting in a nominal degradation of the write energy improvement from 52\% to 51.4\%, on average.\vspace{0.5em} 
}
\section{\textbf{Sensitivity to Granularity}}\label{sens_analysis}
To better understand the interaction between WLC and coset encoding, we analyze the impacts of data block granularity on write energy, the number of updated cells and write disturbance errors. To clarify the difference of reducing one coset candidate and restricting the coset configurability, we also include a 3cosets approach that is as flexible as 4cosets from an encoding perspective but has the same coset candidates as the restricted coset (C1-C3 from Table~\ref{four_coset_candidates}). We also report separately the energy to write the auxiliary and the data symbols.
\subsection{\textbf{Impact of Granularity on Write Energy}}\label{sens-granularity-energy}
Figure~\ref{sens_write_diff_granu} shows the write energy when WLC is used with 4cosets, 3cosets and 3-r-cosets for four data block granularities.
WLCRC-16 (restricted coset with 16-bit block size), achieves the minimum write energy of 6777pJ on average of all the workloads. This is 10\% and 11\% lower than 4cosets and 3cosets, respectively, at their minimum energy point, which is for a data block granularity of 32 bits. 

\begin{figure}[!t]
\centering
\includegraphics[width=0.375\textheight]{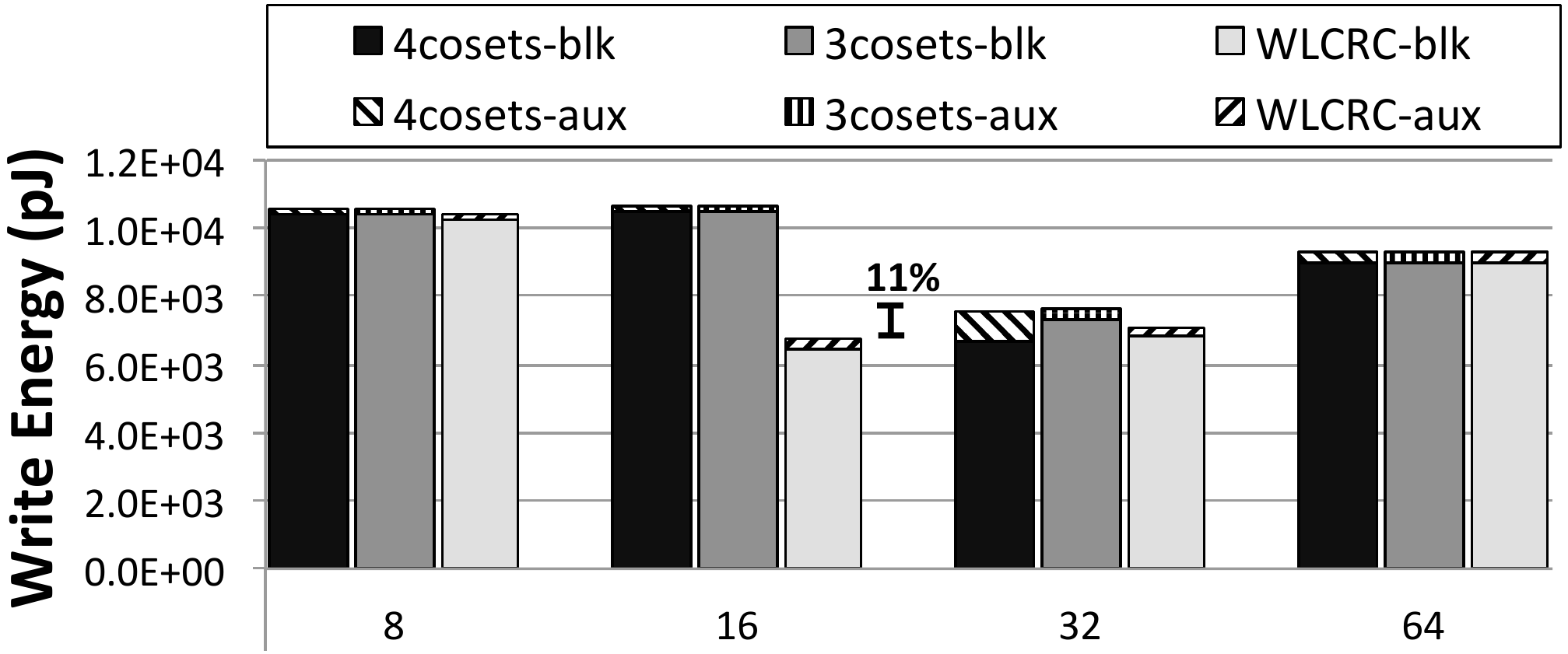}\vspace{-0.5em}
\caption{Write energy comparison for four different data block granularities 8, 16, 32, and 64.
\vspace{-0.5em}
}\label{sens_write_diff_granu}
\end{figure}
To understand why 4cosets and 3cosets require more energy for a 16-bit data block than at a 32-bit block size, as well as more energy than WLCRC at a 16-bit block size, we examine the percentage of compressed memory lines by each scheme. Recall that WLCRC-16 uses one global auxiliary bit per memory line and five auxiliary bits to encode the four independent blocks of each 64-bit word. Thus, WLCRC-16 requires only six bits of compression per word to be applied and this allows for 91\% memory lines to be encoded. In contrast, 4cosets and 3cosets at 16-bit granularity force WLC to provide 8-bits of storage for auxiliary bits in addition to 1 bit for the compressed leading bits, requiring the reclamation of 9-bits. As a result, the percentage of lines that can be encoded drops to 48\%. Of course, since both 4cosets-32 and 3-coset-32 require five bits to store auxiliary bits in the reclaimed part, WLC can be applied on 91\% of lines. This advantage to the application of compression outweighs the encoding advantage of 4cosets-16 and 3cosets-16, respectively, making 32-bit block granularity the minimum energy point for those approaches.

Of course WLCRC-32 is less effective than WLCRC-16 because it can be applied to the same number of memory lines, but has a coarser granularity of encoding that is less flexible for achieving low-energy states.

To further increase the granularity to an 8-bit data block, the reclaimed part must grow to include more than eight bits. Specifically, WLCRC-8 requires seven auxiliary bits to split the word into seven parts noting that the most-significant (eighth) byte will need to be compressed away using WLC to reserve space for the auxiliary bits. When combined with the restricted auxiliary bit, WLC compresses five symbols per word for WLCRC-8, which is only possible for 46\% of memory lines. The flexibility of encoding cannot offset the lost compression effectiveness relative to WLCRC-16. 



Figure~\ref{sens_write_diff_granu} also breaks down the write energy into energy from the auxiliary part and the data part, independently. Note that the average data block energy is reported based the average write energy of data blocks of compressed words + incompressible memory lines. The auxiliary part reaches a maximum of 5.5\% of the total write energy for WLCRC-16, which is less than the 7.8\% of the space the auxiliary part requires. The main reason for the low write energy of the auxiliary part is that the restricted cosets incur less bit changes in the auxiliary part compared to unrestricted cosets. When group cosets switch between C1, C2 and C1, C3 (as shown in Figure~\ref{four_coset_candidates}), the coset candidate C1 which is the most frequent coset, exists in both groups. We allocate the auxiliary bit `0' to the coset candidate C1 that causes the most symbols in the reclaimed part to remain in the low energy states of S1 or S2. For 4cosets, we allocate energy states S1, S2, S3 and S4 to coset candidates C1, C2, C3 and C4. Since coset candidates C1 and C2 are the two most frequent candidates, it keeps the auxiliary part in the low energy states, S1 and S2, for the most of the write requests. 3cosets does not employ C4 (S4) similarly minimizing the high energy states.

\emph{We conclude that the use of WLC to make space for encoding auxiliary bits in the reclaimed part is effective for minimizing write energy. Moreover, the selection of WLCRC-16 is supported as the best trade-off of encoding and block size granularity to minimize write energy.}\vspace{0.25em}
\begin{figure}[!t]
\centering
\includegraphics[width=0.375\textheight]{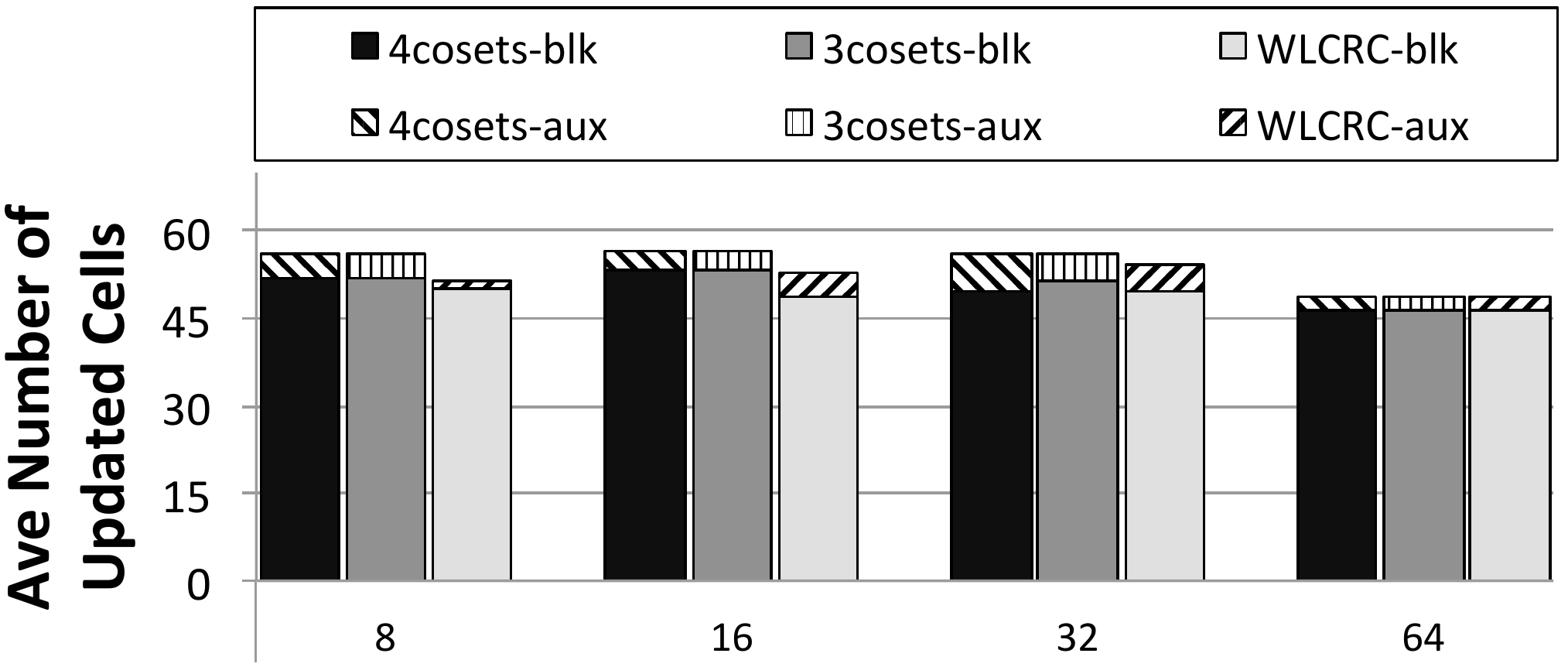}
\vspace{-.1in}
\caption{ Comparison of average updated cells per memory line for four different data block granularities: 8, 16, 32, and 64.
}\label{3_res_cosets_WLC_endurance}
\vspace{0.4em}
\end{figure}
\subsection{\textbf{Impact of Granularity on Endurance}}\label{send-granularity-updated-cells}
Figure~\ref{3_res_cosets_WLC_endurance} shows the number of updated cells (a metric of endurance) as data block granularity scales. At 16-bit granularity, WLCRC reduces the number of updated cells by 8\%, on average compared to WLC+4cosets and WLC+3cosets. In this case, for smaller block granularities (i.e., eight bits) the restricted coset reduces the number of updated cells. For example, at 16-bit granularity, the average number of updated cells in a memory line for WLCRC is 10\% less than for WLC+4cosets and WLC+3cosets, while the auxiliary parts update roughly the same number of cells. As the data block granularity increases to 64, all schemes require similar number of updated cells, which is about 10\% fewer than WLCRC-16. \vspace{0.15em}

\subsection{\textbf{Impact of Granularity on Disturbance}}\label{send-granularity-disturbance}
Figure~\ref{3_res_cosets_WLC_wde} shows the average write disturbance errors for different data block granularities. The average write disturbance errors is approximately three per memory line. However, when data block granularity becomes more coarse, the number of symbol flips decreases, which results in fewer write disturbance errors. One observation from this figure is that the data blocks incur a considerably higher number of write disturbance errors compared to the auxiliary part for WLC-based techniques. Of course this is due to the incidence of 25\% bit flips, on average, of the data block. However, the disturbance errors from the auxiliary bits in the reclaimed part do not change dramatically across different block sizes, as the larger reclaimed parts are only applied when WLC is successful.\vspace{0.25em}
%
%
\section{\textbf{Sensitivity to Energy Levels}}
The analysis in this paper is based on the MLC PCM write energy levels previously reported in the literature~\cite{wang2011energy,bedeschi2009bipolar} shown in Table~\ref{tab:configuration}.
However, subsequent improvements to MLC PCM devices along with better iterative programming approaches may have significantly reduced the energy of writing to intermediate states.
To estimate the effect of these write energy improvements on the effectiveness of WLCRC-16, we 
repeated our experiments with the write energy to high energy states (i.e., S4 and S3) reduced as reported in Figure~\ref{scalability}, while keeping the energy of S1 and S2 unchanged.
The results show that when the write energy cost of these high energy states is reduced by more than 6$\times$, WLCRC-16 still reduces the write energy by 32\% relative to the baseline.
\begin{figure}[!t]
\centering
\includegraphics[width=0.375\textheight]{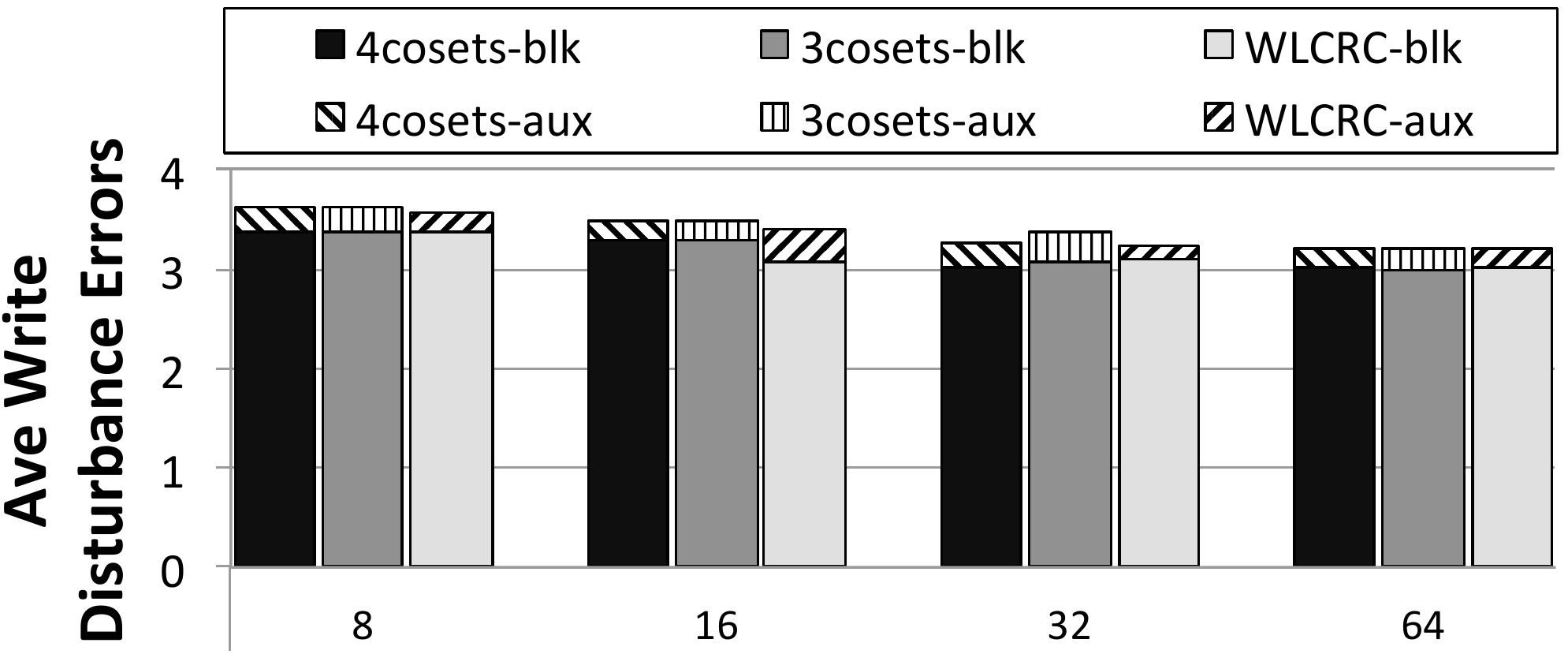}
\vspace{-.05in}
\caption{ Comparison of the write disturbance errors per memory line for four different data block granularities: 8, 16, 32, and 64.\vspace{0.5em}
}\label{3_res_cosets_WLC_wde}

\end{figure}
\begin{figure}[!t]
\centering
\includegraphics[width=0.375\textheight]{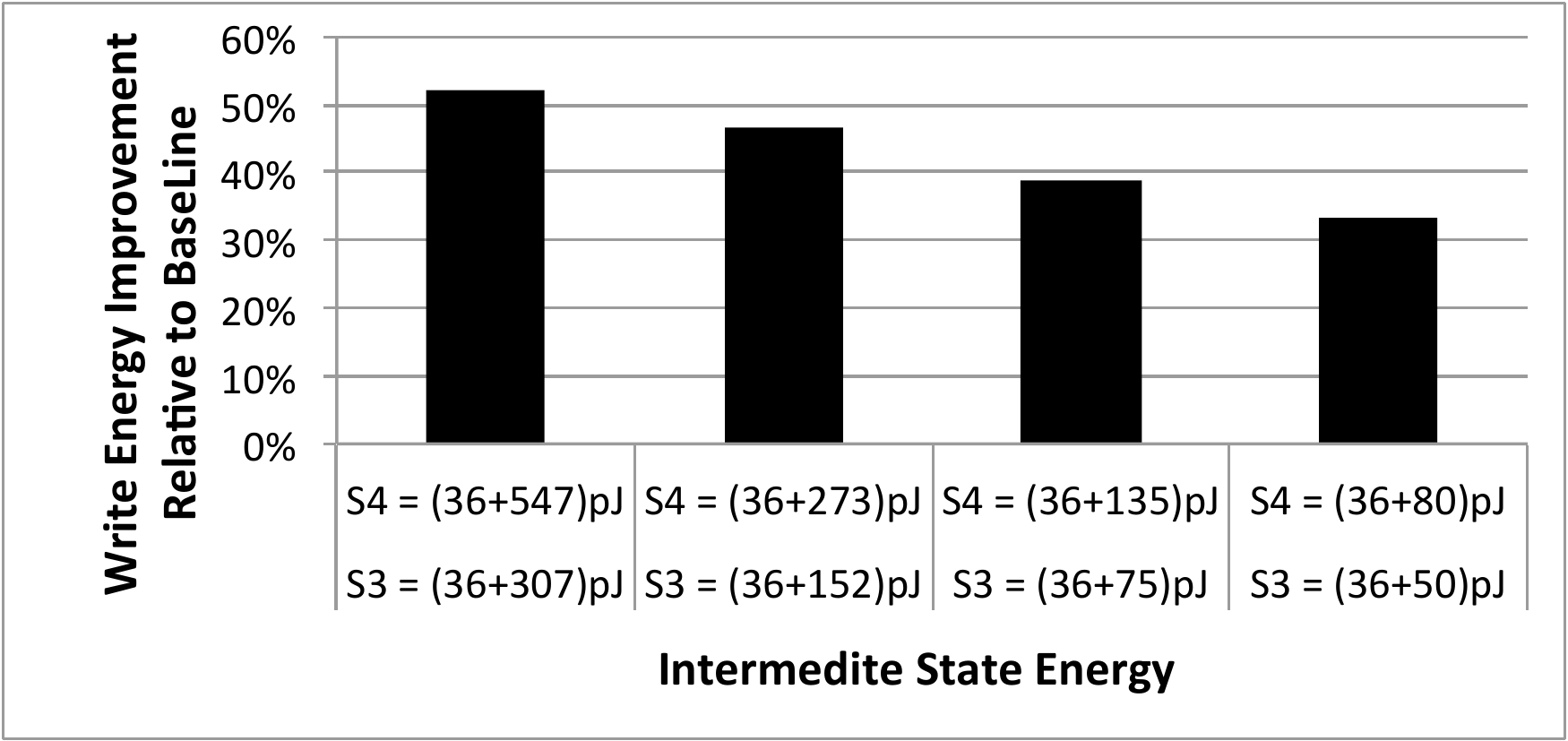}
\caption{ Sensitivity of WLCRC-16 to energy levels.
}\label{scalability}
\end{figure}
\vspace{0.25em}
\section{\textbf{Conclusion}}
The goal of this work is to increase the effectiveness of energy-reduction encoding techniques for PCM by reducing the encoding granularity. We first show that although reducing the encoding granularity reduces the write energy of the data blocks, it considerably increases the auxiliary encoding bits and consequently the energy to write these bits. We thus introduce a new encoding that takes into consideration the bit patterns of workloads to reduce the total PCM write energy.

Specifically, we propose a novel \textit{restricted} coset encoding that largely reduces the number of auxiliary bits compared to known coset encodings while achieving similar write energy reduction. Furthermore, we introduce a new \textit{Word-Level} Compression~(WLC) technique that compresses 91\% of the memory lines while reclaiming enough space in the compressed lines to fit the auxiliary bits.
Finally, we design a new and low hardware overhead architecture, WLCRC, that integrates WLC and restricted coset encoding to effectively reduce the write energy in MLC PCM. Hardware synthesis indicates that WLCRC encoders and decoders incur low area, latency, and energy overheads.

Our experimental results on real workloads show that WLCRC at 16-bit block granularity improves the write energy by about 52\% and 39\%, on average, compared to the baseline and the leading write-minimization approach, respectively. It also improves cell endurance and reliability although no specific provisions are made during the encoding to optimize these metrics. 
In our future work, we plan to extend the WLCRC encoding to be write-disturbance aware, in addition to being write-energy aware. In fact, WLCRC is a general scheme that may be applied to reduce the encoding granularity, whenever coset encoding is relevant.
\section{\textbf{Acknowledgements}}
We thank the anonymous reviewers for their feedback. This work is supported by CS50 merit pre-doctoral fellowship award from the university of Pittsburgh.
\renewcommand\refname{\textbf{References}}
\textbf{\bibliographystyle{reference}}
\bibliography{reference}

\end{document}